# The Spatial Structures in the Austrian COVID-19 Protest Movement: A Virtual and Geospatial Twitter User Network Analysis

Umut Nefta Kanilmaz[a, *], Bernd Resch[a, c] Roland Holzinger[b], Christian Wasner[b], Thomas Steinmaurer[b]

[a]Department of Geoinformatics – Z_GIS, University of Salzburg, Austria;

[b]Department of Communication Science - University of Salzburg, Austria;

[c]Center for Geographic Analysis, Harvard University, Cambridge, MA, USA

Umut Nefta Kanilmaz is a researcher and PhD candidate at the University of Salzburg's Department of Geoinformatics – Z_GIS, focusing on the spatio-temporal analysis of communication networks based on social media data.

Bernd Resch is an Associate Professor at the University of Salzburg's Department of Geoinformatics – Z_GIS and a Visiting Scholar at Harvard University (USA). His research interest revolves around developing machine learning algorithms to analyse human-generated data like social media posts and physiological measurements from wearable sensors.

Roland Holzinger, MA, is a member of the Department of Communication Science at the University of Salzburg researching the conceptual and methodological implications of news trust in a contemporary high-choice media environment.

Christian Wasner, MA, a member of the Department of Communication Science in Salzburg, researches conspiracy theory movements and their networking in the digital space.

Thomas Steinmaurer is a Professor and Head of the Division "Center for ICT&S" at the Department of Communication Studies at the University of Salzburg. The unit is concerned with developments of digital change in society and its significance for individual media practices and democracy on a structural level as well.

*CONTACT Umut Nefta Kanilmaz. Email: umutnefta.kanilmaz@plus.ac.at

# The Spatial Structures in the Austrian COVID-19 Protest Movement: A Virtual and Geospatial Twitter User Network Analysis

The emergence of the COVID-19 pandemic, followed by policy measures to combat the virus, evoked public protest movements world-wide. These movements emerged through virtual social networks as well as local protest gatherings. Prior research has studied such movements solely in the virtual space through social network analysis, thereby disregarding the role of local interaction for protest. This study, however, recognizes the importance of the geo-spatial dimension in protest movements. We therefore introduce a large-scale spatial-social network analysis of a georeferenced Twitter user network to understand the regional connections and transnational influences of the Austrian COVID-19 protest movement through the social network. Our findings reveal that the virtual network is distinctly structured along geographic and linguistic boundaries. We further find that the movement is clearly organized along national protest communities. These results highlight the importance of regional and local influencing factors over the impact of transnational influences for the protest movement.

Keywords: Protest movement, COVID-19, Social media, Social Networks, Geographic Space, Spatial Analysis

**Introduction**

The spread of COVID-19 in 2019 affected the lives of people around the world. National governments in numerous countries tried to prevent a global pandemic by introducing new policy measures, e.g. lockdowns and curfews, oftentimes leading to local protests and demonstrations against them (Kriesi & Oana, 2023). Social media platforms gained popularity for the protest discourse during that time (Jarynowski et al., 2020) and their global reach enabled the spread of (mis-)information across protest participants (Chen et al., 2022; Muric et al., 2021). As a result, emerging 'communities of fate' (Held, 1997) developed a transnational understanding of their political claims giving local COVID-19 protest movements a transnational dimension. Some researchers proclaimed the

'death of geography' (Bates, 1999) in the early days of broad accessibility to the Internet, a theory that would render geospatial distance irrelevant due to closeness in virtual space. Yet during the COVID-19 pandemic, activists still connected digitally and across borders with local, regional, national and international movement partners, organizing protest and negotiating political claims. This created a virtual network structure in social media as well as a spatial structure according to the users' physical locations.

Research around location-based social networks (LBSN) in recent years has already established how methods of spatial network analysis can help gain insight into e.g. cultural differences between regional and interregional communication (Arthur & Williams, 2019). However, most research studying COVID-19 protest movements has focused largely on the social network analysis, disregarding the spatial component (Ahmed et al., 2020; Hung et al., 2020). The combination of both components into a spatial-social network analysis helps understanding which geographic regions are connected through the social network and if connections tend to extend to a transnational degree. This understanding, in turn, brings insight into the geopolitical relationships of digital protest mobilization.

The Austrian COVID-19 protest movement stood out due to its broad and heterogeneous mobilization in the Austrian population (Brunner et al., 2021) and also showed pronounced similarities to the protest movements in the two German-speaking neighbouring countries of Switzerland and Germany (Nachtwey et al., 2020). On the other hand, research suggests that US protest movements originating in right wing groups against the 'plandemic' spread via social media communities to Europe (Fominaya, 2022). In order to understand these regional, national and transnational digital embedding of the Austrian COVID-19 protest movement, our study contributes an extensive spatial-social network analysis for a Twitter user network of protest actors. We structure this analysis along the following research questions:

(1) How does the virtual social network manifest in geographic space?
(2) How are different geographical regions connected through the social network?
(3) How does the framework of socio-spatial network analysis help in understanding the transnationality of the underlying movement?

To answer these research questions, we construct a Twitter user-follower network using a snowball crawling system. Followership between users in Twitter can be modelled as a network graph: Nodes represent Twitter users and the directed edge between them represents the followership. This information about a user's followers, as well as the user's specified location, was retrieved via Twitter's API. By integrating the location information into the user network, we employ established techniques of spatial-social network analysis to gain novel insights into the geospatial manifestation of the Austrian COVID-19 movement. We expect, for instance, a spatially non-uniform distribution of the network nodes and edges. The node distribution is likely to follow the population distribution, resulting in certain regions exhibiting a higher concentration of users and connections. We further assume that regions sharing linguistic similarities will show stronger connections through the virtual network. This phenomenon might be particularly evident among the German-speaking nations of Germany, Austria, Switzerland and Liechtenstein. In terms of the movement's transnational character, we presume a considerable influence of Austria to its neighbouring German-speaking countries. The extent of Austria's impact on the United States represents a particular research aspect.

**Background and Theory**

The potential global reach and connectivity of digital media, combined with the emergence of increasingly global issues, are linked to the development of 'transnational' political online communication (Pfetsch et al., 2019). Transnationality refers to public communication 'that go across borders and transform these borders by establishing structures and cultures of communication that exist beyond the interaction of national states' (Brüggemann & Wessler, 2014). Due to its significance in the digitally mediatized society (Couldry & Hepp, 2018), transnationality currently manifests primarily through issue-related networks of political communication on social media platforms (Pfetsch et al., 2019). Research on transnationality has primarily focused on social movements and important global issues, such as globalization-critical protest movements (Conover et al., 2013), the environmental movement (Hellqvist, 2022), as well as terrorist and far-right movements (Doerr, 2017). Early approaches have tended to emphasize the potential benefits of social media and a shift from the logic of spatial organization to digitally-networked activism (Sorce & Dumitrica, 2022).

Although digital media play a significant role in the development of a transnational discourse, evidence increasingly suggests that political online communication also remains heavily localized in geographic space. Shared language is an important prerequisite for interactions between social media users, including for political debates and mobilization (Bastos et al., 2013; Takhteyev et al., 2012). Online networks also occur significantly along shared national identities and cultural backgrounds (Arthur & Williams, 2019; Casero-Ripollés et al., 2020). When social movements emerge from online networks, the intertwining with analog forms of protest on the ground (Eis, 2021) and local aspects (Hopke, 2016) are of great importance. National states often remain the main target for social movements, as they usually represent the immediate responsibility for global problems (Brunnengräber, 2012). At the local level, Casero-Ripollés et al (2020) have shown that users in densely populated cities and in geographic proximity to political centres of power are more actively engaged in political debates. Digital communication networks of social movements are therefore fundamentally localized in spatial contexts, but at the same time contain a transnational potential, meaning, the potential for actors in different localities connect digitally and develop a cross-border awareness of political demands. The question of transnationality is therefore tied to the actors who create it (Sorce & Dumitrica, 2022).

**Related Work**

This study examines a virtual online network of Twitter users and its manifestation in geographic space, using methods of spatial and social network analysis. The following section will therefore summarize related work in both areas.

Ahmed et al. (2020) study the spread of a specific conspiracy theory using social network analysis on keyword-filtered Twitter data. Their focus lies on determining central users by calculating their betweenness centrality scores, revealing users that actively engaged in sharing conspiracy theories.

Rauchfleisch & Kaiser (2020) present a study that analyses a YouTube network of conspiracy-related channels focusing on a far-right political spectrum. They use a snowball-style crawling of the platform's recommender system, where first a list of YouTube channels with politically far-right content is collected and recommended channels are retrieved from this initial start list. The channels were connected with a directed edge if a channel was recommended by another. Filtering for significant edges using the method of (Serrano et al., 2009) and running a clustering algorithm, the authors were able to identify distinguishable communities.

A methodologically similar study was done for the German Querdenken movement deriving network data from Telegram (Zehring & Domahidi, 2023), identifying an influence of QAnon and far right communities.

**All of the mentioned studies above focus on the analysis of the virtual online social network, disregarding the network's spatial component.** Scellato et al. (2010) analyse a Twitter user network with methods of spatial network analysis. To describe each network's manifestation in geographic space, they introduce the concept of node locality and a geographic clustering coefficient. The former quantifies a node's local embedding in contrast to global interactions, and the latter introduces a weighing based on node distance to the known clustering coefficient (Boccaletti et al., 2006). The authors find that Twitter users are more likely to interact with users that are spread globally instead of users that are at closer distances.

Following up in (Scellato et al., 2011), the authors further investigate the average geographic distance between users and the probability of a link between network nodes as a function of distance. The results are compared to two different statistical null models: one keeping the geographical location and shuffling the links, and the other one vice versa. A notable finding is that the occurring triangles, i.e., a direct connection between all three nodes, are spread on a wide geographical scale and that the length of a link does not affect the probability of belonging to a social triangle.

Another study looking into the spatial properties of a Twitter network focuses on comparing social and spatial distance of users within the US (Stephens & Poorthuis, 2015). For each user in the network, an ego-alter subnetwork is created, and its transitivity and density calculated. The results show that with increasing distance between the ego and its alters, the network density and transitivity decreases, indicating that the closer the ego is to its alters, the higher the social clustering is. Furthermore, networks with a standard distance of less than 500 km have high transitivity, which decreases with physical distance to a constant level.

Although much research has been performed on social networks structures, to the best of our knowledge, no publication studies the Austrian COVID-19 by comparing the social and spatial network structures.

**Methods of Spatial-social Network Analysis**

*Overall Workflow*

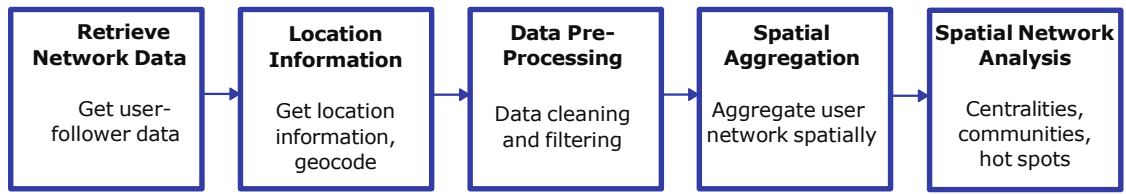

Figure 1. The workflow followed in this study.

This study aims to understand the Austrian COVID-19 protest movement by means of spatial-social network analysis using the workflow displayed in Figure 1. The network data of user-follower relationships as well as each user's location was retrieved using Twitter's API. After geocoding and preprocessing the data, we aggregated the user network spatially and calculated network centralities, applied a community detection algorithm and performed a spatial hot spot analysis. The following section will describe the single steps in more detail.

*Data Retrieval*

We retrieved the network data of user-follower relationships using Twitter's API with a snowball crawling procedure, which started crawling the follower network from a selected set of start or seed users. A meaningful selection of seed users is important especially with regard to the transnationality analysis (Sorce & Dumitrica, 2022), as those users determine the network's overall structure and ensure the relevance to the protest context. For this analysis, potential seed users were identified through extensive qualitative research of Austrian Corona protest media coverage. The selected protest actors had to have a Twitter account and had to be located in Austria based their profile information. Their active engagement into the protest was ensured through manual inspection of their Twitter accounts, specifically active posting and content re-tweeting behaviour as well as the amount of followers. Furthermore, seed users had to reflect heterogeneity of the protest actors with regard to their ideological background and functional role (Borbáth, 2023), as well as their status as opinion leader on twitter in terms of followers. The selected five seed users were characterized by

- right-wing and anti-system ideological backgrounds
- functional roles of politicians, experts and citizens
- and different numbers of followers (few <500, medium 500-5,000, many >5,000).

We retrieved the network data of user-follower relationships using Twitter's API with a snowball crawling procedure. More than one iteration of followers needed to be crawled to be able to observe a connected network structure instead of several unconnected trees. Crawling more iterations, however, might weaken the effects of interests in the network that are due to the selected seed users. For example, current literature reports a value for the degree of separation as 4.59 in Twitter networks (Watanabe & Suzumura, 2013). After retrieving the followers of these initial actors, the network was constructed. Users were modelled as network nodes and directed edges were added from a user to their followers.

*Location Information and Geocoding*

In this study, we assume that a user's location is the one specified in their location profile. This information can be retrieved using Twitter's API and then be passed to a geocoding service, which translates the textual information into geographic coordinates. OpenStreetMaps's Nominatim was used for this purpose. All user nodes that did not specify a location or had locations that could not be geocoded were removed from the network. The resulting georeferenced network consisted of approximately 800,000 nodes distributed over 160,000 unique locations and with 1.2 million connecting edges. Table 1 summarizes the network structure.

Table 1. Number of users, edges, and locations in the network before and after filtering for location and geocoding.

| Characteristics | Absolute Number (%) |
| --- | --- |
| **Users** | |
| Unique Users | 1609811 (100 %) |
| With location information in profile | 884227 (54,93 %) |
| Geocodable location information | 786541 (48,86 %) |
| **Edges** | |
| Unique directed edges | 4325470 (100 %) |
| Edges between geocodable nodes | 1182474 (27,34 %) |
| **Locations** | |
| Unique locations | 244239 (100 %) |
| Locations with geocodable reference | 159 506 (65,31 %) |

The amount left after filtering for location was approximately 49 % of all user nodes and 28 % of all edges. Less than one percent of the user nodes were geocoded into areas that have a population count of zero according to the world population dataset provided by (Center For International Earth Science Information Network-CIESIN-Columbia

University, 2018). These user locations, e.g., 'Antarctica' and 'Bir Tawil', appeared to be fake information upon manual inspection of the user profiles, and nodes with such locations were dropped. Furthermore, 13 % of all users provided country or continent names as location information, which were geocoded as the centroid of the countries or continent's bounding box. As this location information was too imprecise for analysing regional connectivity, those nodes were removed as well.

*Spatial Aggregation of User Network Data*

One goal within this study is to understand the connectivity of geographical regions through the virtual user network. This can be achieved by defining geographical regions in the form of a regular grid, where each cell represents a region, and then aggregating the point network onto the grid layer, using the following procedure:

- The number of users falling into each grid cell was counted and assigned as node weight to the grid cell in the resulting aggregated network.
- Two cells in the grid network were connected with a directed edge between them if there was at least one user-follower relationship with a user and follower located in the respective cells.
- The total number of directed connections between two grid cells was counted. This number was assigned as an edge weight in the resulting grid network.

This procedure results in a network of connected grid cells, which allows to study the connections of regions through the underlying user network. The grid size was chosen according to the research question. For instance, **understanding the transnationality** of the movement introduced the concept of nations into the analysis. We therefore aggregated the user network into the administrative borders of the world's countries. The question about the **connectivity of regions** through the user network was approached by aggregating the network nodes on a global hexagonal grid with a cell size of 80,000 km. Choosing an evenly sized grid, as opposed to clustering points into variable cluster sizes, was important in this analysis as we aimed to calculate several network centralities and compare their magnitude and their spatial distribution. Using hexagons was beneficial visualizing the results. The choice of cell size was of importance as it impacted the results of the evaluation based on the spatial aggregation (Wong, 2004). The size of a region for the world-wide analysis was chosen by visual analysis considering that a single cell should not cover whole or several European countries, as we studied an Austrian protest movement and wanted to identify regional connections. Analysing the properties of this network indicated **areas of interest** (AOI). Those areas could then be further examined by filtering the user network geographically for user nodes and edges that fall into the AOI and performing the spatial network analysis for Europe on an even finer grid with a cell size of 100 square kilometres.

*Spatial Network Analysis*

The aggregation in the previous steps resulted in a network of connected regions. We calculated four different centrality metrics to indicate regions of importance in the grid cell network.

The number of edges connecting a node is called the **degree centrality** (Freeman, 1978) and reveals its importance in terms of the number of connections it has. As we investigate a Twitter user network, it is important to note that the user-follower relationships are not symmetric. When a user follows another, they retrieve information from the other's feed, but not vice versa. This means that the edge direction has to be regarded and in- and outgoing edges have to be counted and evaluated separately. A node with an *out-degree* of N represents *a user who is followed by N other users*, whereas an

*in-degree* of N means that *this node follows N other users* and retrieves their feed. In other words, the direction of the edge represents the direction in which information travels in the network.

Determining how close a node is to other nodes in the network can be done with the **closeness centrality** (Freeman, 1978; Wasserman & Faust, 1994). Network nodes with a high closeness centrality value have the shortest path length to other nodes in the network and are therefore 'closer'. For the grid network this means that cells or regions that have a high closeness centrality are able to spread information through the network more efficiently. The **betweenness centrality** (Brandes, 2001; Freeman, 1977) indicates the importance of a network node with regard to information flow: nodes with a high betweenness centrality lie more frequently on the shortest path that connects two other nodes in the network. In the aggregated network, cells with a high betweenness centrality can be interpreted as regions with a high amount of control over the information flow.

**Community detection** algorithms detect clusters of nodes in the network such that nodes within the group are more densely connected within each other than with the rest of the network. Communities in the grid cell network can be regarded as geographical regions that are stronger connected to each other than to other regions. There are many possible methods for detecting communities and we used the Louvain algorithm (Blondel et al., 2008). This method optimizes on the network's modularity and is particularly known to be a fast algorithm even on large networks (Mothe et al., 2017).

For the calculated centralities in the grid cell network, we perform a **hot spot analysis** using the Getis-Ord Gi* method (Getis & Ord, 1992) to detect spatial clusters of central regions. These clusters are important as they mark regions that are central for the underlying user network of COVID-19 protest actors.

**Results**

The results of the analysis are presented in this section in a descriptive fashion, whereas a thorough interpretation of the results will follow in the Discussion section.

*Spatial Properties of the Twitter User Network*

Figure 2 shows a visualization of the network using a so-called edge bundling algorithm (Ersoy et al., 2011) where edges have been bundled into 'highways' of connections based on similar positional information. The line width indicates the number of edges that follow along a connecting route. We can see that the users in the network are distributed over most populated parts of the world and many connections lie within Europe, within the US and also between those two continents.

In Figure 3, we can see that the normalized geodesic edge length distribution using a bin size of 115 km, calculated using the Freedman-Diaconis rule. The distribution is right-skewed, peaks at around 350 km and shows a second, significantly smaller increase of samples between 7,000 and 8,000 km. The dip between 3,000 km and 6,000 km delineates a 'ring' around Europe defined by the Atlantic and Indian Oceans.

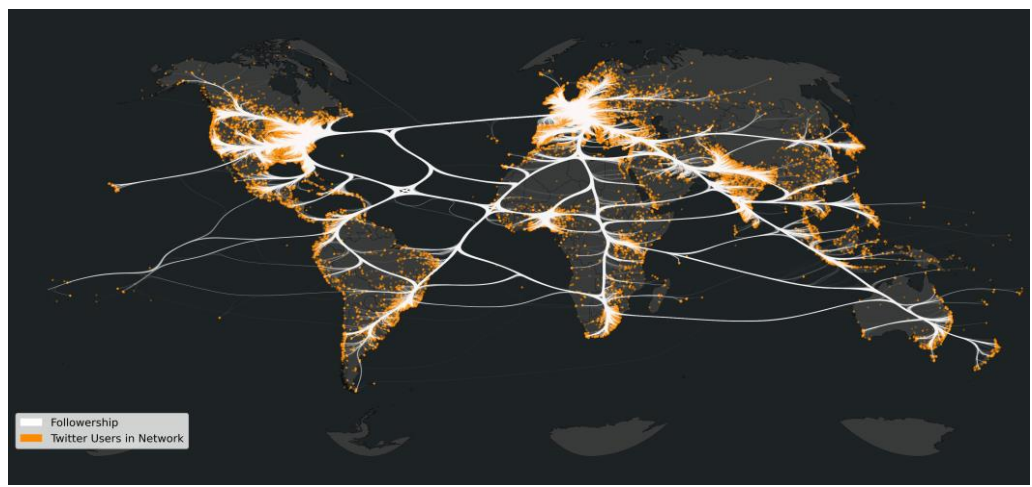

Figure 2. Edge-bundled visualization of the georeferenced Twitter user network of the Austrian COVID-19 protest movement.

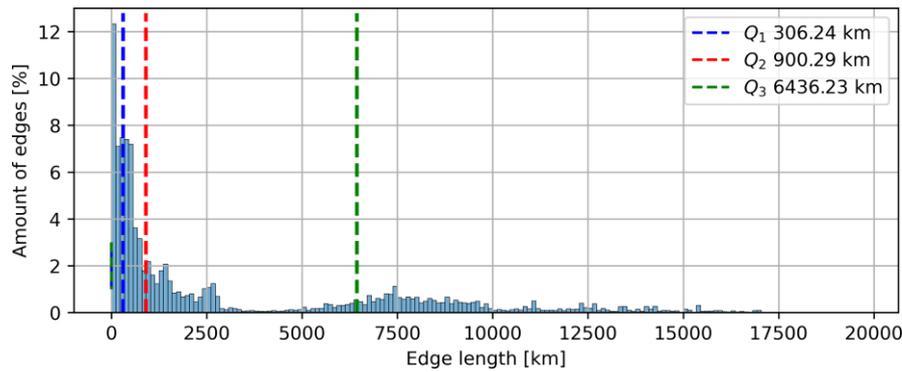

Figure 3. Normalized distribution of network edge lengths in bins of 115 km.

*Spatial Aggregation on Country Level*

A chord diagram visualizes the connections between countries through the network in Figure 4. For reasons of clarity, only the 15 most frequently occurring edges which make up 67 % of all edges were visualized, finding that 44 % of all edges occur within the same nations, namely Germany (22.98 %), the US (11.71 %) and Austria (9.41%). 4.73 % of edges lead from Austria to Germany, 3.21 % from Germany to the US and 1.71 % from the UK to the US.

Next, we examine the edges originating in Austria, which indicates the countries with followers of Austrian users and, in terms of transnationality, which nations Austria potentially influences. Table 2 summarizes the total number of edges in percent that originate in Austria. The majority of edges originating in Austria (46.9 %) are inner-Austrian edges, followed by Germany with 23.65 % and the US with 7.27 %. Other countries, like the UK, Switzerland, and France each have less than a 3 % share of the edges that start in Austria. When looking into the number normalized by the country's population, we find that the inner-Austrian edges dominate even more strongly and make up 60.96 % of edges. Smaller countries like Liechtenstein, Switzerland and Luxembourg become more apparent compared to the absolute count of edges. Notably, the amount of edges going out to Germany is significantly smaller when looking into the normalized count (3.28 % normalized compared to 23.63 % absolute).

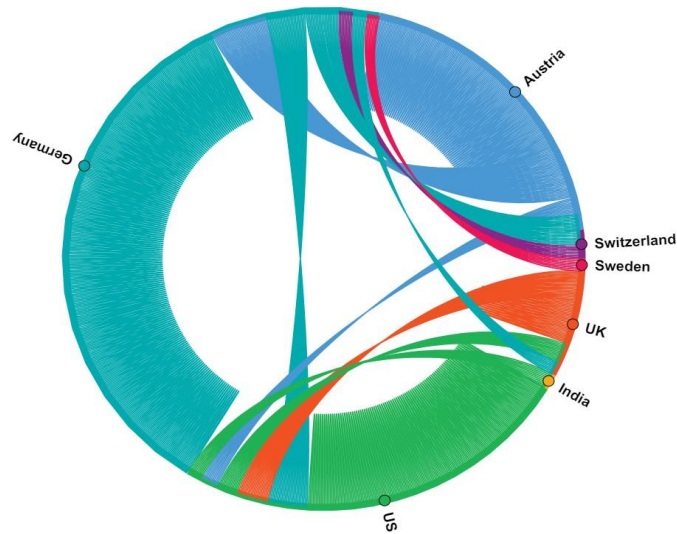

Figure 4. Chord diagram of absolute number of edges in the country level network

Table 2. Destination countries of edges out of Austria ordered by relative amount in % (left) and normalized by country population (right).

| Absolute Numbers | | Normalized by Population | |
|---|---|---|---|
| Country | (%) | Country | (%) |
| Austria | 46.89 | Austria | 60.96 |
| Germany | 23.63 | Other | 14.66 |
| Other | 11.73 | Liechtenstein | 11.21 |
| US | 7.27 | Germany | 3.28 |
| UK | 2.92 | Switzerland | 2.99 |
| Switzerland | 2.22 | Luxembourg | 1.83 |
| France | 1.52 | Gambia | 1.26 |
| Canada | 1.38 | Monaco | 1.14 |
| Netherlands | 0.82 | Andorra | 0.96 |
| Spain | 0.81 | Ireland | 0.93 |
| Italy | 0.81 | Iceland | 0.78 |

*Spatial Aggregation on a Global Grid*

*Network Centralities*

We calculate the in- and out-degree, closeness and betweenness centralities of the grid cell network next. Cells with a high centrality indicate regions that have a high potential of influence over the information flow in the network. The distributions of the calculated centralities are shown in Figure 5. Betweenness and out-degree, have heavily right-skewed distributions where 87.72 % of cells have a value of zero.

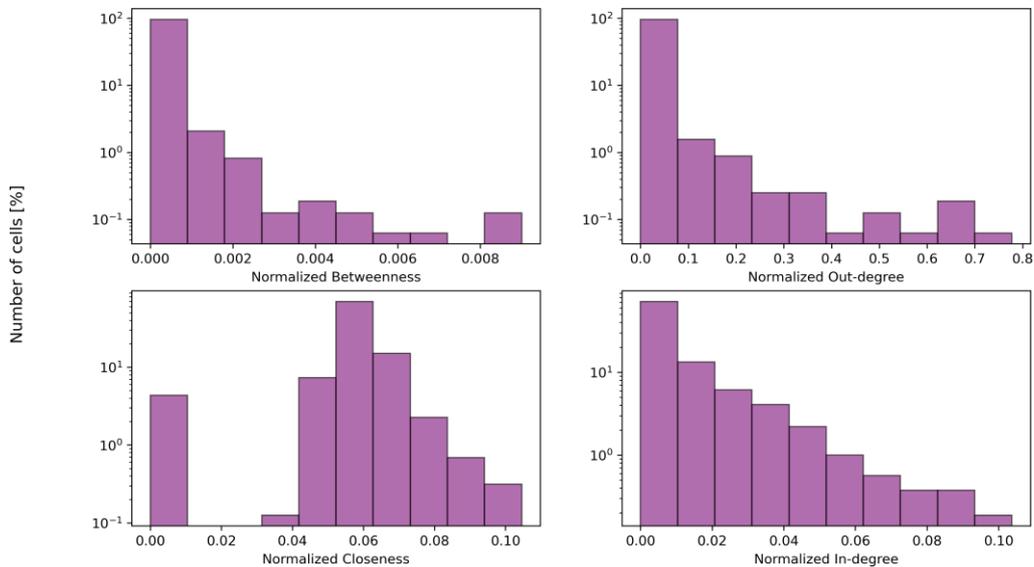

Figure 5. Value distribution of each centrality measure.

As the value distributions are not the same and not normally distributed, we use Spearman's rank coefficient to test for a monotonic relationship between those four centralities. The resulting correlation matrix is visualized in Figure 6. It shows that betweenness and out-degree, as well as closeness and in-degree are strongly related with a correlation of 99.9 and 98.9 %, respectively.

Knowing that the mentioned centrality pairs form a monotonic relationship in virtual space, the question arises if high or low values of the centralities also coincide in the same grid cell. We therefore cluster the centrality distributions into three value ranges: all cells with a centrality value of zero are assigned to the lowest bin. This is especially important for out-degree and betweenness centrality, as the majority of cells have a value of 0. For the remaining values, we calculate the median and assign the lower half of the distribution to the middle cluster and the upper half to the high cluster. Those low, middle and high clusters of the two centralities are then visualized using a bivariate map as shown in Figure 7, where regions of high-high and low-low clusters become visible. For betweenness and out-degree centralities, we found a spatial cluster of cells falling into the high-high bins in central Europe. Single cells in the high-high value range could also be found in the US, especially on the East Coast. Most cells overall were low in both betweenness and out-degree. The closeness and in-degree centralities showed far more cells that fell into the high-high value bins, distributed along the more populated parts of the world: covering almost the entirety of the US and Europe, as well as parts of India and the coasts of South America and Africa.

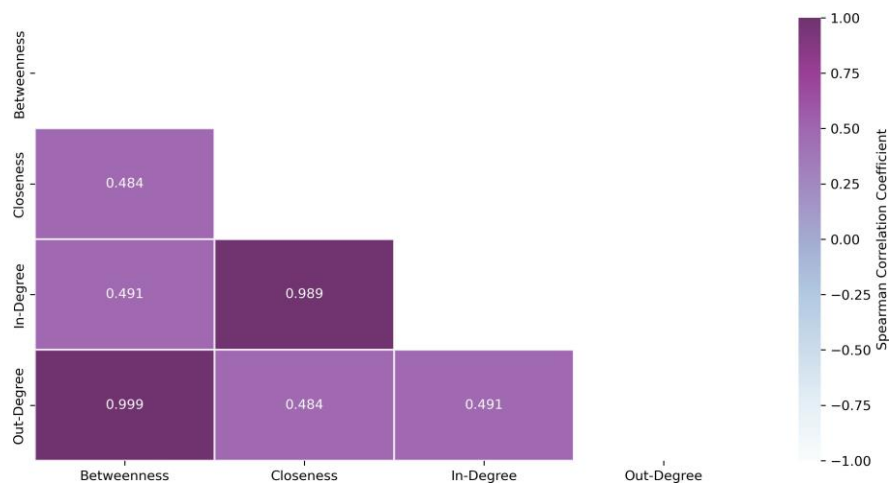

Figure 6. Spearman correlation matrix of the in-degree, out-degree, closeness and betweenness centralities for the grid cell network.

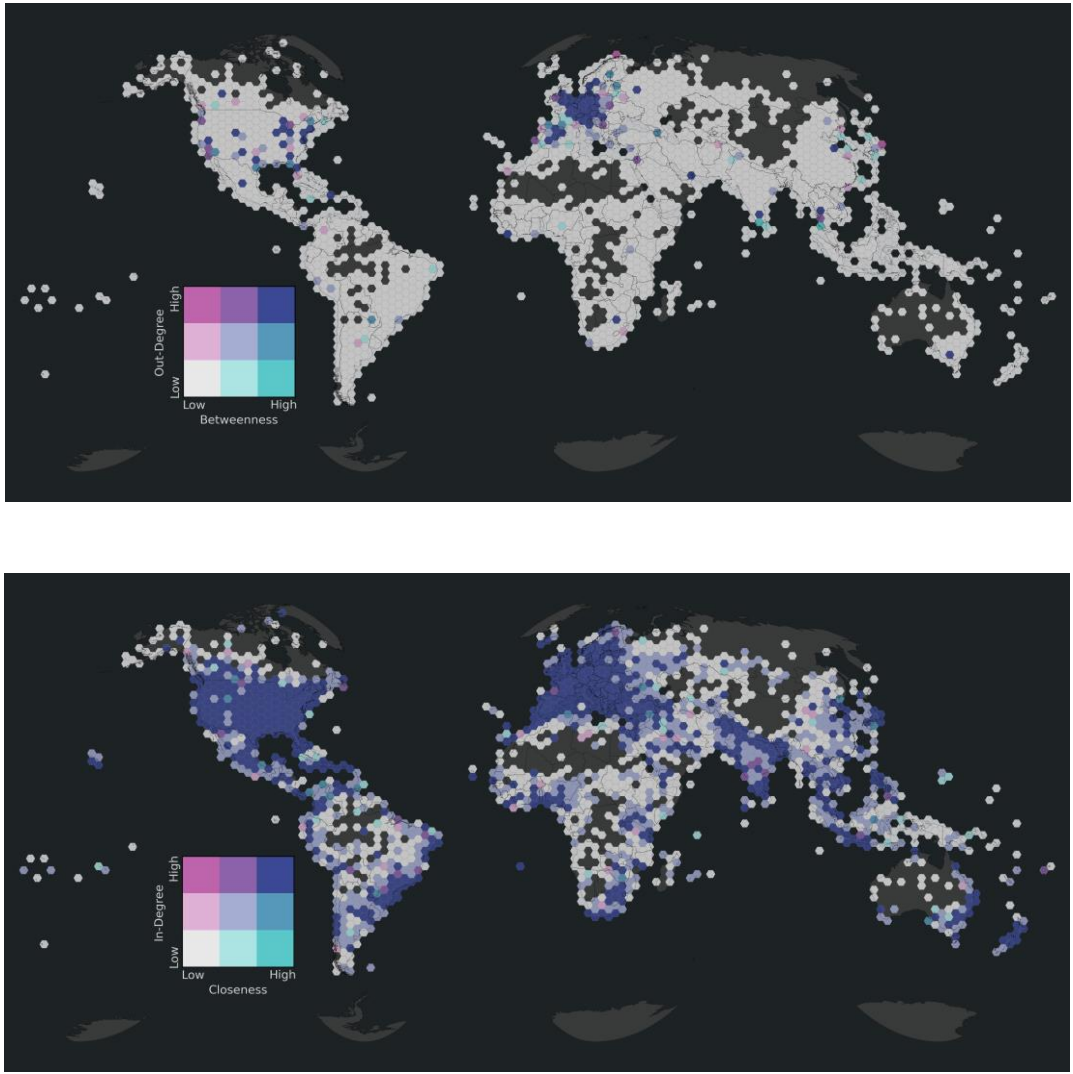

Figure 7. Bivariate choropleth map of betweenness and out-degree (upper), as well as closeness and in-degree (lower)

*Community Detection*

Figure 8 illustrates the 10 largest communities detected within the grid cell network, with each grid cell color-coded to represent its assigned community as determined by the algorithm. The most extensive community, consisting of 833 cells and shaded in dark blue, forms a global presence encompassing English-speaking regions, like the US, the UK, India and large parts of Africa. The second-largest community, coloured in orange, consists of 223 cells and has a noticeable presence in Russia, the South and South-East of Brazil and also includes Eastern Europe. Community number three is rooted in central

Europe, with single cells scattered across the globe. It is also noticeable that some communities redraw the geographical boundaries of countries: Mexico and Turkey are clearly distinguishable in brown and light green, respectively. There are also several communities that lack a clear regional concentration, e.g. communities number four, five, and seven. Communities number nine and ten are the smallest communities consisting of 11 cells and partly cover Spain and Italy, respectively.

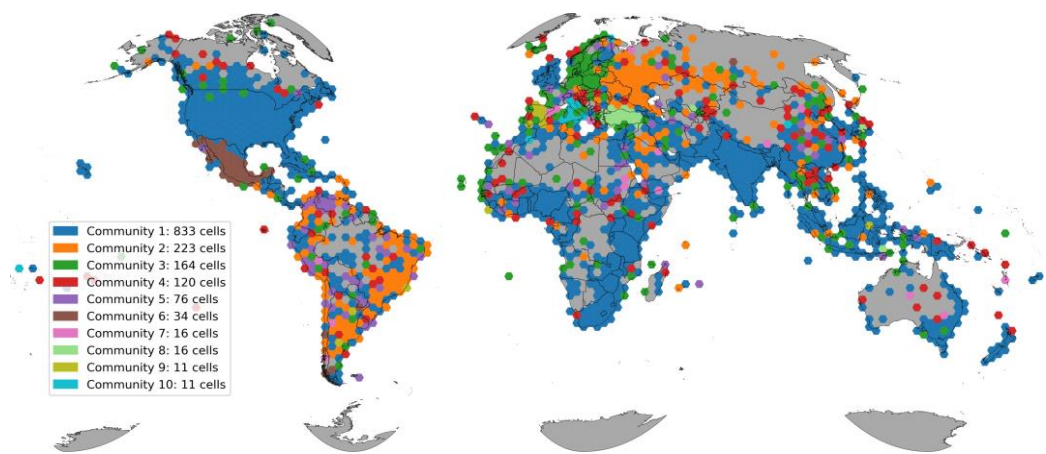

Figure 8. 10 largest communities in the grid cell network detected by the Louvain algorithm.

*Analysing the European Subnetwork*

*Community Detection*

The presented previous results indicate that with regard to the betweenness and out-degree centralities, the European continent and the US are the main AOIs. However, the results of community detection suggest that the US and the DACH region do not form a dedicated community, instead, the US seems to be embedded into a large community structure spanning all over the world. In order to identify important regions for the Austrian COVID-19 protest movement in more detail, we analysed the European subnetwork. Therefore, we filtered for nodes and edges in the user network that were within a bounding box surrounding Europe. We then aggregated the subnetwork to a hexagonal grid of 100 square kilometre area size and re-calculated communities, as well as closeness and betweenness centralities on the finer grid.

Figure 9 presents the outcomes of community detection within the European subnetwork and Figure 10 the according distribution of the community cells across the countries. Some communities are spread over several countries, whereas the others show a clearly dominant presence in a distinct country. Specifically, the largest community, coloured in blue, covers the area of the United Kingdom and is scattered across France, Italy and Spain. Community number two forms a distinct cluster centred in Germany with more than 50 % of cells located there, and less than 10 % being dispersed throughout other European countries Communities number three is strongly rooted in Austria, coloured in orange in the map in Figure 9, with 20 % percent of all cells being located there, as well as France, Italy and Spain. The fourth-largest community is dominant in Switzerland and reaches into Germany and France. The remaining six communities are notably smaller in terms of the number of cells belonging to them, making up less than 10 % of cells compared to the largest cluster. Those communities exhibit clear associations with country borders: Netherlands, Turkey, Switzerland, the Balkan countries, Denmark, and Spain are each associated with significantly smaller, but very distinct communities. Remarkably, the Catalan region around Barcelona stands out as a dedicated community, distinguishing itself from the broader Spanish context.

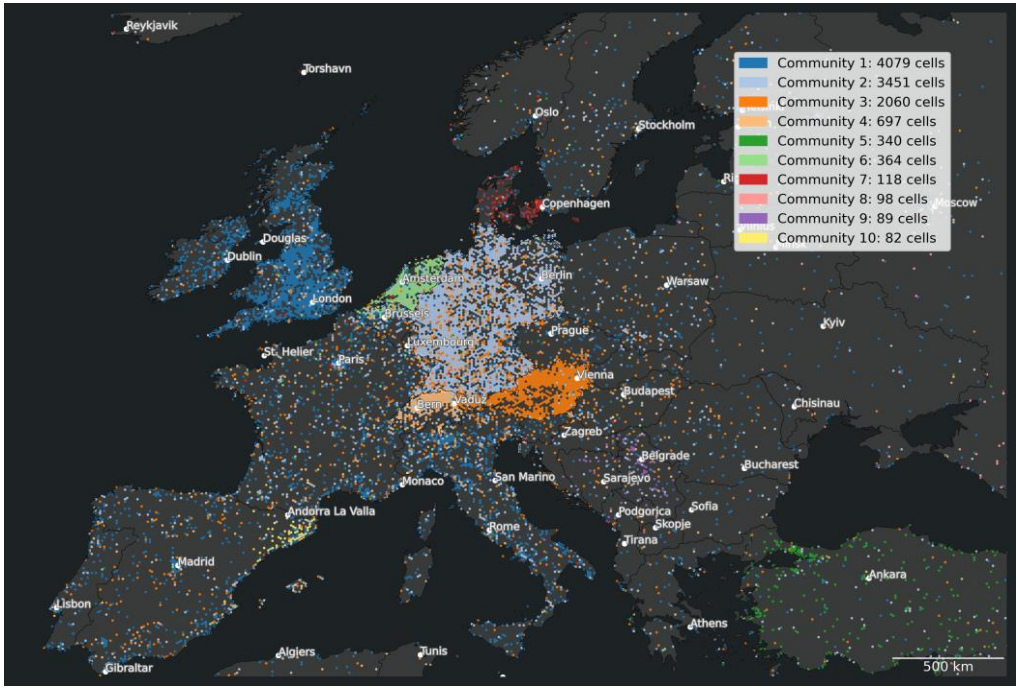

Figure 9. Community detection for the European subnetwork.

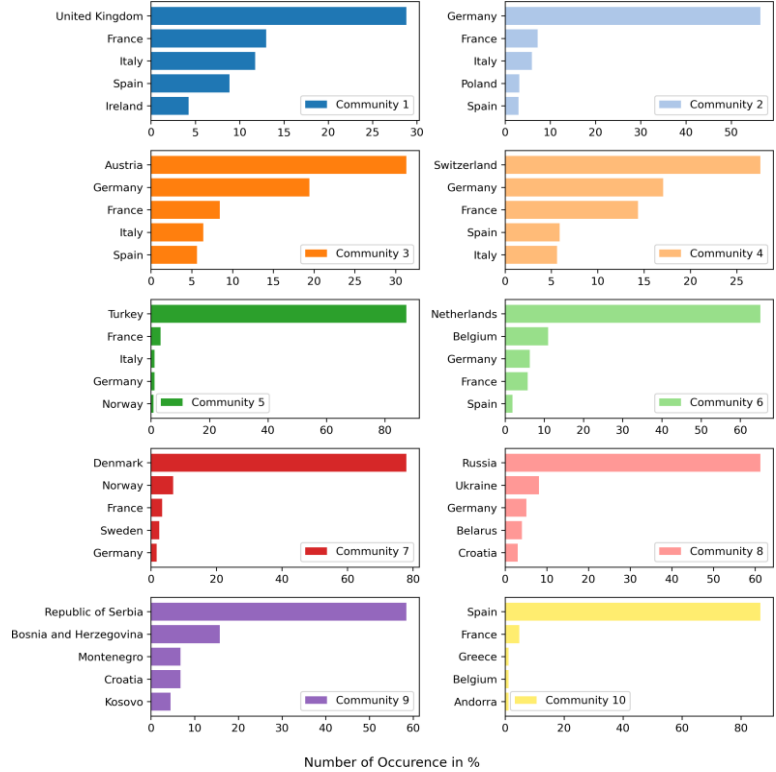

Figure 10. Distribution of communities per country.

*Hot Spot Analysis of Centralities*

In section 5.3, we already identified areas of interest with regard to high centrality values on a coarse grid. This analysis can be extended further with spatial hot spot analysis of the European subnetwork's centralities. These hot spots indicate regional clusters of users who were important for the information exchange in the virtual protest movement network. A hot spot analysis using the Getis-Ord GI* algorithm with 30 nearest neighbours as conceptualization of spatial relationship was performed on the aggregated European subnetwork for betweenness and closeness centrality and the results are shown in Figure 11.

Closeness centrality hot spots could be found in Germany, the German-speaking part of Switzerland and parts of Austria, especially Vienna, Graz and the area of upper Austria. The cells in the UK were mostly non-significant, with the exceptions of the greater London area, which stands out as a hot spot with strong significance as well as weaker hotspots in the area of Manchester. Less significant hot spots can be found in Paris and Amsterdam. Ireland, Denmark and Serbia, however, were dominated by cold spots.

The hot spot map of the betweenness centrality is clearly dominated by cells classified as statistically not significant. There were clusters around urban areas, however, that were categorized as hot spots. Strong hot spots with 99 % confidence could be found in the urban areas of Vienna, Berlin and London and in Munich. Less significant hot spots with 90 to 95 % confidence were detected in the south of Stuttgart, as well as in the area around Essen.

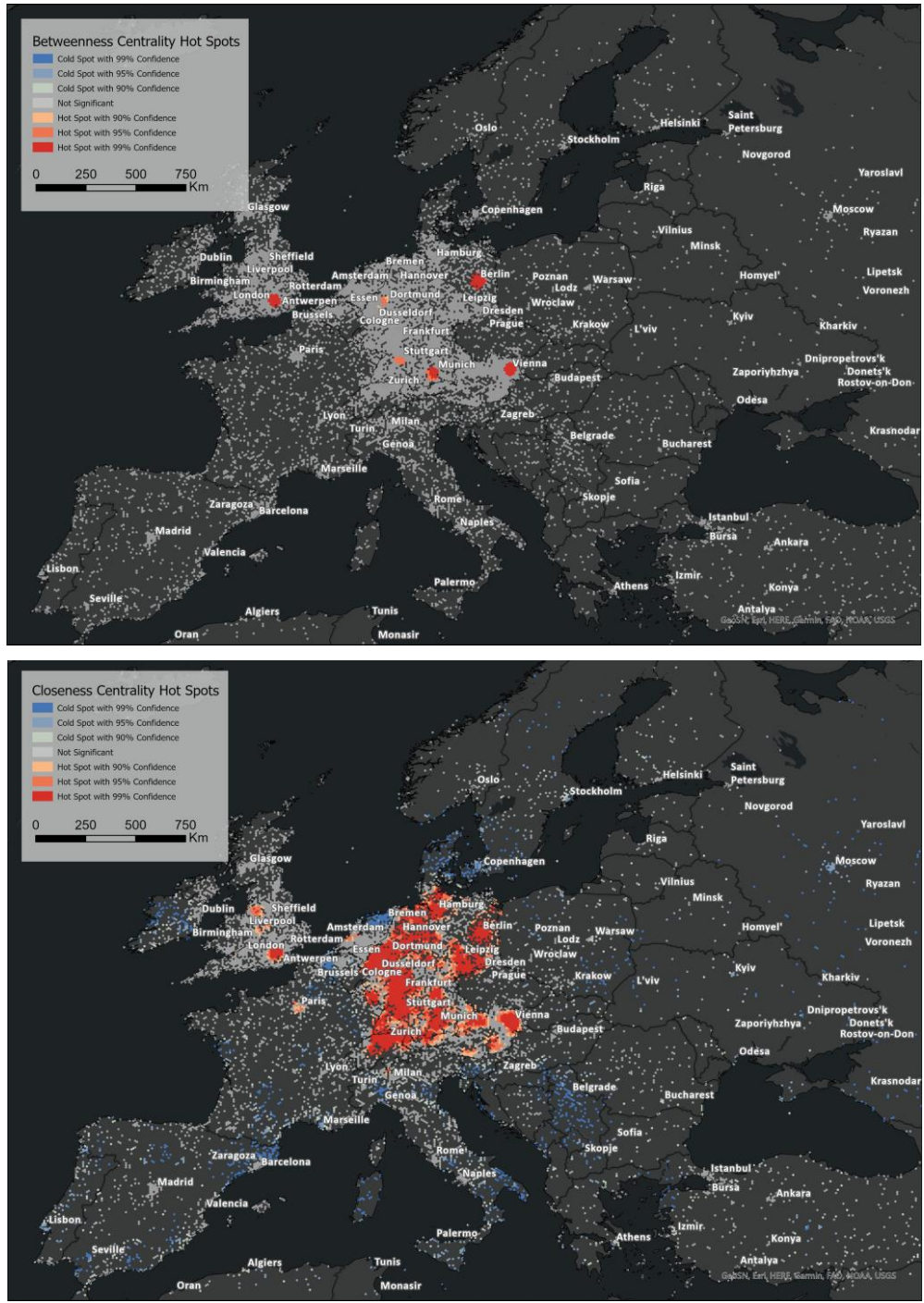

Figure 11. Hot spot maps of closeness (upper) and betweenness centrality (lower).

**Discussion**

This section discusses the methods and interprets the results obtained by the spatial-social network analysis. In answering the research questions in section 1, we demonstrate how analysing the social network of protest actors helps further understanding the underlying protest movement.

*Interpretation of Results*

*How does the virtual social network manifest in geographic space?*

When examining the overall network plot, it is evident that **the node distribution partly resembles the world population distribution**. A high concentration of nodes can be found in the coastal areas of the United States, Europe, India and Brazil with fewer nodes in e.g. Russia, and China. The distribution of points, as well as the geodesic distribution of edge lengths closely mirrors the spatial nature of the network. Almost 50 % of all edges have a length up to 900 km, which suggests that **most edges in the network appear within continents**.

The discovered communities on the globally aggregated network indicate a **strong role of languages** in structuring the network. Notably, the US and several countries, mostly English-speaking, are integrated into one large community. This suggests that the numerous connections observed in and to the US may be attributed to the overall large network, the shared language, and the strong use of social media in the U.S., rather than an effect that is due to the initial network context of the Austrian COVID-19 protest movement.

The results of community detection within the European subnetwork further underline these findings, where the English-speaking UK is the centre of the largest community. The high number of community cells in other European countries and urban centres could suggest a language-based followership. The significance of a shared language can also be observed in a community covering the Slavic-speaking region in Serbia and Bosnia. In addition to this apparent linguistic structure in the network, the detected communities are **clearly organized along country boundaries**, as can be seen in the main communities in the Netherlands and Denmark. For the DACH region, we observe that Austria, Germany, and Switzerland each predominantly belong to one network community that also follows the respective country boundaries. However, a scattering of community cells across the other German speaking countries can also be overserved. **These results suggest that the online political movement is structured in primarily national protest communities through the Austrian protest actor network.** This conclusion is supported by the fact that during the COVID-19 pandemic, nation states increasingly initiated independent crisis management measures, which then became the target of protest movements with a strong national orientation (Kriesi & Oana, 2023).

*How strongly are different geographical regions connected through the social network?*

In terms of closeness centrality hot spots, again the DACH region stands out, with the highest number of hot spots. Interestingly, while we were able to detect several distinct communities, it is **only within the DACH region that we see a well-connected region** that plays a crucial role in the dissemination of information, as measured by network centrality. Furthermore, the findings of both betweenness as well as closeness centrality hot spots analysis **emphasize the importance of urban areas**. The Austrian hot spots around Vienna, Graz and Linz, for example, not only represent the country's largest cities, but were also regions of increased protest activities (*Annäherung von Rechtsextremen und Coronaleugnern*, 2023). These results indicate that in particular urban regions are more strongly interconnected through the Austrian protest actor network, indicating a structure along an addition urban-rural dimension. This finding is consistent with existing research, identifying more active participation in online political debates especially by users in densely populated cities (Casero-Ripollés et al., 2020).

*How does the framework of socio-spatial network analysis help in understanding the transnationality of the underlying movement?*

Considering that the study subject is an Austrian protest movement, the high absolute number of inner-Austrian connections is expected, as is the dominance of Germany as a highly populated German-speaking country, together with the US, which has the highest rate of social media use world-wide. However, the number of edges normalized by population of the countries emphasizes inner-Austrian edges even more and lessens the effect of German edges significantly. In line with this, community detection within the European subnetwork finds a distinct community centred on Austria that at the same time extends considerably to the DACH region and in particular Germany.

In summary, we observe **a potential for transnationality for the Austrian protest network in the DACH region** which appears to be supported by the shared language. With regard to online social networks, the shared language is a known prerequisite for interactions between social media users, especially with regard to political debates and mobilization (Bastos et al., 2013; Brunnengräber, 2012). Our findings are further supported by, firstly, literature that observes pronounced similarities between the Corona protest movements of Switzerland and Germany, where actors network and mobilize across borders (Nachtwey et al., 2020), secondly, the prevailing national structure of the network and lastly the absence of a clear cross-country user community on Austrian territory. This interpretation is in line with existing literature finding that the main addressees for social movements often remain nation states (Brunnengräber, 2012).

*Discussion of the Methodology*

The network data was retrieved when starting the analysis in late July 2022, which is not the peak time of political protest in Austria. It was not possible to retrieve follower relationships for that time due to Twitter's API limitations. The retrieved network data was still to be relevant for the protest, all of the selected set of seed actors were actively engaging in the protest discourse on Twitter at the time of data retrieval.

The user locations within the network were approximated as the information provided in their profiles. It is important to acknowledge the limitations of this approach, as the provided locations references cannot be verified for actual presence. However, this approximation is sufficient for the research aims, which are not concerned with the precise position of users but rather in understanding the broader geo-spatial connectivity through the virtual network.

Missing location information and unsuccessful or imprecise geocoding resulted in the removal of 52 % percent of nodes and 73 % of all edges. Despite extensive filtering, significant structures in both the virtual and spatial network were detected, as demonstrated through the results of community detection and hot spot analysis.

The methods chosen for network analysis in this study are considered state-of-the-art within the field and therefore appropriate for the goals of this analysis – providing considerable analytical insight rather than methodological advancement.

The presented analysis assumes that the Austrian COVID protest plays a significant role for the networking among the interconnected users. This assumption can be evaluated in future research by investigating the connection between geographical localization, virtual networks, and exchanged content.

## Conclusion

This study employed spatial-social network analysis to the Twitter network of Austrian COVID-19 protest actors. By examining the interconnections of geographic regions through the virtual social network, we gained further insights into the underlying movement and sought to address a gap in existing literature. Our findings revealed that the network representing the protest movement is dominated by strong national clusters that also form along linguistic boundaries. On a finer geographic scale, hot spot analysis revealed the importance of urban centres for the movement. We further find that Austria's potential for influence is weak among other nations and the inner-Austrian connections appears dominant, outweighing even the potential of influence into Germany, a country that speaks the same language and is ten times the size of Austria in terms of population. This underlines the importance of the local and regional factors in the Austrian protest movement over the potential significance of shared-language. Further research can extend this work to examine the content spread through the network of protest actors and study it's spatio-temporal evolution.

## Funding Details

This work was funded by the State of Salzburg.

## Disclosure statement

The authors report there are no competing interests to declare.

## Data availability statement

The data that support the findings of this study are available from the corresponding author upon reasonable request.

**References**

Ahmed, W., Vidal-Aluball, J., Downing, J., & Seguí, F. L. (2020). COVID-19 and the 5G Conspiracy Theory: Social Network Analysis of Twitter Data. *Journal of Medical Internet Research*, *22*(5), e19458. https://doi.org/10.2196/19458

*Annäherung von Rechtsextremen und Coronaleugnern*. (2023, July 8). https://www.kleinezeitung.at/politik/innenpolitik/6305206/Oberoesterreich_Annaeherung-von-Rechtsextremen-und-Coronaleugnern

Arthur, R., & Williams, H. T. P. (2019). The human geography of Twitter: Quantifying regional identity and inter-region communication in England and Wales. *PLOS ONE*, *14*(4), e0214466. https://doi.org/10.1371/journal.pone.0214466

Bastos, M. T., Raimundo, R. L. G., & Travitzki, R. (2013). Gatekeeping Twitter: Message diffusion in political hashtags. *Media, Culture & Society*, *35*(2), 260–270. https://doi.org/10.1177/0163443712467594

Bates, S. (1999). The Death of Geography, the Rise of Anonymity and the Internet. *International Information Communication and Education*, *18*(2), 261–265.

Blondel, V. D., Guillaume, J.-L., Lambiotte, R., & Lefebvre, E. (2008). Fast unfolding of communities in large networks. *Journal of Statistical Mechanics: Theory and Experiment*, *2008*(10), P10008. https://doi.org/10.1088/1742-5468/2008/10/P10008

Boccaletti, S., Latora, V., Moreno, Y., Chavez, M., & Hwang, D.-U. (2006). Complex networks: Structure and dynamics. *Physics Reports*, *424*(4), 175–308. https://doi.org/10.1016/j.physrep.2005.10.009

Borbáth, E. (2023). Differentiation in Protest Politics: Participation by Political Insiders and Outsiders. *Political Behavior*. https://doi.org/10.1007/s11109-022-09846-7

Brandes, U. (2001). A faster algorithm for betweenness centrality. *The Journal of Mathematical Sociology*, *25*(2), 163–177. https://doi.org/10.1080/0022250X.2001.9990249

Brüggemann, M., & Wessler, H. (2014). Transnational Communication as Deliberation, Ritual, and Strategy. *Communication Theory*, *24*(4), 394–414. https://doi.org/10.1111/comt.12046

Brunnengräber, A. (2012). Ein neuer Bewegungszyklus: Von der NGOisierung zur Occupy-Bewegung. *Forschungsjournal Soziale Bewegungen*, *25*(1), 42–50. https://doi.org/10.1515/fjsb-2012-0106

Brunner, M., Daniel, A., Knasmüller, F., Maile, F., Schadauer, A., & Stern, V. (2021). *Corona-Protest-Report. Narrative – Motive – Einstellungen*. SocArXiv. https://doi.org/10.31235/osf.io/25qb3

Casero-Ripollés, A., Micó-Sanz, J.-L., & Díez-Bosch, M. (2020). Digital Public Sphere and Geography: The Influence of Physical Location on Twitter's Political Conversation. *Media and Communication*, *8*(4), 96–106. https://doi.org/10.17645/mac.v8i4.3145

Center For International Earth Science Information Network-CIESIN-Columbia University. (2018). *Gridded Population of the World, Version 4 (GPWv4): Population Count, Revision 11* [dataset]. Palisades, NY: NASA Socioeconomic Data and Applications Center (SEDAC). https://doi.org/10.7927/H4JW8BX5

Chen, E., Jiang, J., Chang, H.-C. H., Muric, G., & Ferrara, E. (2022). Charting the Information and Misinformation Landscape to Characterize Misinfodemics on Social Media: COVID-19 Infodemiology Study at a Planetary Scale. *JMIR Infodemiology*, *2*(1), e32378. https://doi.org/10.2196/32378

Conover, M. D., Ferrara, E., Menczer, F., & Flammini, A. (2013). The Digital Evolution of Occupy Wall Street. *PLOS ONE*, *8*(5), e64679. https://doi.org/10.1371/journal.pone.0064679

Couldry, N., & Hepp, A. (2018). *The mediated construction of reality*. John Wiley & Sons. https://books.google.com/books?hl=en&lr=&id=yJ9RDwAAQBAJ&oi=fnd&pg=PR3&dq=The+mediated+construction+of+reality+hepp&ots=ByouDNc0Ij&sig=TT4UEBx_grnFciuZ8PaigiEByP0

Doerr, N. (2017). Bridging language barriers, bonding against immigrants: A visual case study of transnational network publics created by far-right activists in Europe. *Discourse & Society*, *28*(1), 3–23. https://doi.org/10.1177/0957926516676689

Eis, A. (2021). Digitale Kommunikation und transnationale Öffentlichkeit(en). In M. S. Hubacher & M. Waldis (Eds.), *Politische Bildung für die digitale Öffentlichkeit: Umgang mit politischer Information und Kommunikation in digitalen Räumen* (pp. 109–130). Springer Fachmedien. https://doi.org/10.1007/978-3-658-33255-6_6

Ersoy, O., Hurter, C., Paulovich, F., Cantareiro, G., & Telea, A. (2011). Skeleton-Based Edge Bundling for Graph Visualization. *IEEE Transactions on Visualization and Computer Graphics*, *17*(12), 2364–2373. https://doi.org/10.1109/TVCG.2011.233

Fominaya, C. F. (2022). Mobilizing During the Covid-19 Pandemic: From Democratic Innovation to the Political Weaponization of Disinformation. *American Behavioral Scientist*, 00027642221132178. https://doi.org/10.1177/00027642221132178

Freeman, L. C. (1977). A Set of Measures of Centrality Based on Betweenness. *Sociometry*, *40*(1), 35–41. https://doi.org/10.2307/3033543

Freeman, L. C. (1978). Centrality in social networks conceptual clarification. *Social Networks*, *1*(3), 215–239. https://doi.org/10.1016/0378-8733(78)90021-7

Getis, A., & Ord, J. K. (1992). The Analysis of Spatial Association by Use of Distance Statistics. *Geographical Analysis*, *24*(3), 189–206. https://doi.org/10.1111/j.1538-4632.1992.tb00261.x

Held, D. (1997). Democracy and Globalization. *Global Governance*, *3*(3), 251–267.

Hellqvist, E. (2022). *Land Under Attack! : A Study of the Framing Process of the Transnational Social Movement Fridays for Future*. https://urn.kb.se/resolve?urn=urn:nbn:se:uu:diva-494194

Hopke, J. E. (2016). Translocal anti-fracking activism: An exploration of network structure and tie content. *Environmental Communication*, *10*(3), 380–394. https://doi.org/10.1080/17524032.2016.1147474

Hung, M., Lauren, E., Hon, E. S., Birmingham, W. C., Xu, J., Su, S., Hon, S. D., Park, J., Dang, P., & Lipsky, M. S. (2020). Social Network Analysis of COVID-19 Sentiments: Application of Artificial Intelligence. *Journal of Medical Internet Research*, *22*(8), e22590. https://doi.org/10.2196/22590

Jarynowski, A., Semenov, A., & Belik, V. (2020). Protest Perspective Against COVID-19 Risk Mitigation Strategies on the German Internet. In S. Chellappan, K.-K. R. Choo, & N. Phan (Eds.), *Computational Data and Social Networks* (pp. 524–535). Springer International Publishing. https://doi.org/10.1007/978-3-030-66046-8_43

Kriesi, H., & Oana, I.-E. (2023). Protest in unlikely times: Dynamics of collective mobilization in Europe during the COVID-19 crisis. *Journal of European Public Policy*, *30*(4), 740–765. https://doi.org/10.1080/13501763.2022.2140819

Mothe, J., Mkhitaryan, K., & Haroutunian, M. (2017). Community detection: Comparison of state of the art algorithms. *2017 Computer Science and Information Technologies (CSIT)*, 125–129. https://doi.org/10.1109/CSITechnol.2017.8312155

Muric, G., Wu, Y., & Ferrara, E. (2021). COVID-19 Vaccine Hesitancy on Social Media: Building a Public Twitter Data Set of Antivaccine Content, Vaccine Misinformation, and Conspiracies. *JMIR Public Health and Surveillance*, *7*(11), e30642. https://doi.org/10.2196/30642

Nachtwey, O., Frei, N., & Schäfer, R. (2020). *Politische Soziologie der Corona-Proteste* [Working Paper]. Universität Basel. https://doi.org/10.31235/osf.io/zyp3f

Pfetsch, B., Heft, A., & Knüpfer, C. (2019). Transnationale Öffentlichkeiten in der Digitalen Gesellschaft: Konzepte und Forschungsperspektiven. In *Politik in der digitalen Gesellschaft* (pp. 83–102). transcript Verlag. https://doi.org/10.1515/9783839448649-005

Rauchfleisch, A., & Kaiser, J. (2020). The German Far-right on YouTube: An Analysis of User Overlap and User Comments. *Journal of Broadcasting & Electronic Media*, *64*(3), 373–396. https://doi.org/10.1080/08838151.2020.1799690

Scellato, S., Mascolo, C., Musolesi, M., & Latora, V. (2010). Distance matters: Geo-social metrics for online social networks. *Proceedings of the 3rd Wonference on Online Social Networks*, 8.

Scellato, S., Noulas, A., Lambiotte, R., & Mascolo, C. (2011). Socio-Spatial Properties of Online Location-Based Social Networks. *Proceedings of the International AAAI Conference on Web and Social Media*, *5*(1), Article 1. https://doi.org/10.1609/icwsm.v5i1.14094

Serrano, M. Á., Boguñá, M., & Vespignani, A. (2009). Extracting the multiscale backbone of complex weighted networks. *Proceedings of the National Academy of Sciences*, *106*(16), 6483–6488. https://doi.org/10.1073/pnas.0808904106

Sorce, G., & Dumitrica, D. (2022). Transnational dimensions in digital activism and protest. *Review of Communication*, *22*(3), 157–174. https://doi.org/10.1080/15358593.2022.2107877

Stephens, M., & Poorthuis, A. (2015). Follow thy neighbor: Connecting the social and the spatial networks on Twitter. *Computers, Environment and Urban Systems*, *53*, 87–95. https://doi.org/10.1016/j.compenvurbsys.2014.07.002

Takhteyev, Y., Gruzd, A., & Wellman, B. (2012). Geography of Twitter networks. *Social Networks*, *34*(1), 73–81. https://doi.org/10.1016/j.socnet.2011.05.006

Wasserman, S., & Faust, K. (1994). *Social Network Analysis: Methods and Applications*. Cambridge University Press. https://doi.org/10.1017/CBO9780511815478

Watanabe, M., & Suzumura, T. (2013). How social network is evolving? A preliminary study on billion-scale twitter network. *Proceedings of the 22nd International Conference on World Wide Web*, 531–534. https://doi.org/10.1145/2487788.2487988

Wong, D. W. S. (2004). The Modifiable Areal Unit Problem (MAUP). In D. G. Janelle, B. Warf, & K. Hansen (Eds.), *WorldMinds: Geographical Perspectives on 100 Problems: Commemorating the 100th Anniversary of the Association of*

*American Geographers 1904–2004* (pp. 571–575). Springer Netherlands. https://doi.org/10.1007/978-1-4020-2352-1_93

Zehring, M., & Domahidi, E. (2023). German Corona Protest Mobilizers on Telegram and Their Relations to the Far Right: A Network and Topic Analysis. *Social Media + Society*, *9*(1), 20563051231155106. https://doi.org/10.1177/20563051231155106

**Captions**

- Figure 1. The workflow followed in this study
- Figure 2. Edge-bundled visualization of the georeferenced Twitter user network of the Austrian COVID-19 protest movement.
- Figure 3. Normalized distribution of network edge lengths in bins of 115 km.
- Figure 4. Chord diagram of absolute number of edges in the country level network
- Figure 5. Value distribution of each centrality measure.
- Figure 6. Spearman correlation matrix of the in-degree, out-degree, closeness and betweenness centralities for the grid cell network.
- Figure 7. Bivariate choropleth map of betweenness and out-degree (upper), as well as closeness and in-degree.
- Figure 8. 10 largest communities in the grid cell network detected by the Louvain algorithm.
- Figure 9. Community detection for the European subnetwork.
- Figure 10. Distribution of communities per country.
- Figure 11. Hot spot maps of closeness (upper) and betweenness centrality (lower).